\documentclass[10pt,conference]{IEEEtran}

\usepackage{cite}

\usepackage[utf8]{inputenc}

\usepackage{url}

\usepackage{csquotes}

\usepackage{graphicx}
\graphicspath{{images/}}

\usepackage{multirow}

\usepackage{makecell}

\usepackage{array}

\usepackage{textcomp}
\def\BibTeX{{\rm B\kern-.05em{\sc i\kern-.025em b}\kern-.08em T\kern-.1667em\lower.7ex\hbox{E}\kern-.125emX}}

\begin{document}

\title{Assuring the Evolvability of Microservices: Insights into Industry Practices and Challenges}

\author{
    \IEEEauthorblockN{
        Justus Bogner\IEEEauthorrefmark{1}\IEEEauthorrefmark{2},
	    Jonas Fritzsch\IEEEauthorrefmark{2}\IEEEauthorrefmark{1},
		Stefan Wagner\IEEEauthorrefmark{2},
		Alfred Zimmermann\IEEEauthorrefmark{1}
	}
	\IEEEauthorblockA{
    	\IEEEauthorrefmark{1} University of Applied Sciences Reutlingen, Germany\\
    	\IEEEauthorrefmark{2} University of Stuttgart, Germany\\
    	\{justus.bogner, alfred.zimmermann\}@reutlingen-university.de\\ 
    	\{jonas.fritzsch, stefan.wagner\}@informatik.uni-stuttgart.de
    }
}

\maketitle

\thispagestyle{plain}
\pagestyle{plain}

\begin{abstract}
While Microservices promise several beneficial characteristics for sustainable long-term software evolution, little empirical research covers what concrete activities industry applies for the evolvability assurance of Microservices and how technical debt is handled in such systems. Since insights into the current state of practice are very important for researchers, we performed a qualitative interview study to explore applied evolvability assurance processes, the usage of tools, metrics, and patterns, as well as participants' reflections on the topic. In 17 semi-structured interviews, we discussed 14 different Microservice-based systems with software professionals from 10 companies and how the sustainable evolution of these systems was ensured. Interview transcripts were analyzed with a detailed coding system and the constant comparison method.

We found that especially systems for external customers relied on central governance for the assurance. Participants saw guidelines like architectural principles as important to ensure a base consistency for evolvability. Interviewees also valued manual activities like code review or boy scouting, even though automation and tool support was described as very important. Source code quality was the primary target for the usage of tools and metrics. Despite most reported issues being related to Architectural Technical Debt (ATD), our participants did not apply any architectural or service-oriented tools and metrics. While participants generally saw their Microservices as evolvable, service cutting and finding an appropriate service granularity with low coupling and high cohesion were reported as challenging. Future Microservices research in the areas of evolution and technical debt should take these findings and industry sentiments into account.
\end{abstract}


\section{Introduction}
Fast moving markets and the age of digitization require that software can be quickly adapted or extended with new features. If change implementations frequently happen under time pressure, the sustainable evolution of a long-living software system can be significantly hindered by the intentional or unintentional accrual of technical debt \cite{Avgeriou2016}. The quality attribute associated with software evolution is referred to as \textit{evolvability} \cite{Rowe1998}\cite{Bode2009}: the degree of effectiveness and efficiency with which a system can be adapted or extended. Evolvability is especially important for software with frequently changing requirements, e.g. internet-based systems.

To provide sufficient confidence that such a system can be sustainably evolved, software professionals apply a set of numerous activities that we refer to as \textit{evolvability assurance}. Activities are usually either of an analytical nature to identify issues or of a constructive nature to remediate issues \cite{Wagner2013}. This includes for example techniques like code review or refactoring, standardization, guidelines, conscious technical debt management, and the usage of tools, metrics, or patterns. For larger systems, these activities often form a communicated assurance process and are an important part of the development workflow.

Microservices constitute an important architectural style that prioritizes evolvability \cite{Newman:15:MS}. A key idea here is that fine-grained and loosely coupled services that are independently deployable should be easy to change and to replace. Moreover, one of the postulated Microservices characteristics is \textit{evolutionary design} \cite{Fowler2015}. While these properties provide a nice theoretical basis for evolvable systems, they offer no concrete and universally applicable solutions. As with each architectural style, the implementation of a concrete Microservice-based system can be of arbitrary quality. The \enquote{service cutting} activity has been described as especially challenging and several approaches have been proposed by academia to support it \cite{Fritzsch2019}. Apart from this, very little scientific research has covered the areas maintenance, evolution, or technical debt for Microservices. Examples include the tracking and management of Microservices dependencies \cite{Esparrachiari2018}, a service evolution modeling technique for Microservices \cite{Sampaio2017}, and the applicability of service-based maintainability metrics for Microservices \cite{Bogner2017}.

In addition to this sparse scientific state of the art, there are also very few empirical studies on the industry state of practice. Very little is known about what evolvability assurance processes and techniques companies use for Microservices or if these are different compared to other architectural styles. In the general area of service-based systems, Schermann et al. \cite{Schermann2016} even describe a mismatch between what academia assumes and what industry actually does. An analysis of industry practices in this regard could identify common challenges, showcase successful processes, and highlight gaps and deficiencies. This would also provide insights into how industry perceives academic approaches specifically designed for service orientation, e.g. service-oriented maintainability metrics. Results of such a study could help to design new and more suited evolvability assurance processes or techniques.

We therefore conducted detailed interviews with 17 Germany-based software professionals from 10 different companies. They described 14 different systems with various Microservices characteristics and their concrete evolvability assurance process including tool, metric, and pattern usage. We also talked with them about the evolution qualities of Microservices, how Microservices influence the assurance process, and their perceived challenges for evolvability in the context of Microservices.


\section{Related Work}
Several empirical studies about challenges for the adoption of Microservices exist. Baškarada et al. \cite{Baskarada2018} conducted 19 in-depth interviews with experienced architects. They discussed opportunities and challenges associated with the adoption and implementation of Microservices. Four types of corporate systems with different levels of suitability for Microservices were identified. Especially large corporate systems of record like Enterprise Resource Planning (ERP) were not seen as appropriate targets. In this context, organizational challenges such as DevOps methodologies would be less serious for Information and Communication Technology enterprises than for traditional organizations, where IT was perceived as a \enquote{necessary evil} \cite{Baskarada2018}.

Ghofrani and Lübke \cite{Ghofrani2018} conducted a similar empirical survey among 25 practitioners that were mainly developers and architects. Their objective was to find perceived challenges in designing, developing, and maintaining Microservice-based systems. The results reveal a lack of notations, methods, and frameworks for architecting Microservices. Several participants named the distributed architecture as responsible for the challenging development and debugging of the system. Participants generally prioritized optimizations in security, response time, and performance over aspects like resilience, reliability, and fault tolerance.

In their interviews with 10 Microservices experts from industry, Haselböck et al. \cite{Haselbock2018} focused on design areas of Microservices and associated challenges. The study identified 20 design areas and their importance as rated by the participants. Design principles and common challenges from earlier mapping studies could be confirmed by the authors. Similar to our qualitative study, interviewees' rationales are discussed as well. Microservices design is a fundamental aspect of their evolvability later on. As such, the study can be seen as a valuable contribution to the topic at hand.

Some studies also focus on evolvability-related areas like technical debt and architecture smells. Carrasco et al. \cite{Carrasco2018} conducted a literature review to gather migration and architecture smells. The authors derived best practices, success stories, and pitfalls. By digesting 58 different sources from academia and grey literature, they presented nine common bad smells with proposed solutions. This study extends the work of Taibi and Lenarduzzi \cite{Taibi2018} who defined 11 bad smells by analyzing 265 bad practices experienced by 72 developers from 61 companies. The results of their two-year interview phase were mapped onto a list of main pitfalls extracted from grey literature and practitioner talks. Out of 16 extracted pitfalls, only six were confirmed by the interviewees. Wrong service cuts during the split of a monolith turned out to be the most critical challenge, resulting in potential maintenance issues later on.

The same authors \cite{Lenarduzzi2018} investigate technical debt interest by means of a long term case study where they monitored the migration of a monolithic legacy system to Microservices. The study aims to characterize technical debt and its growth comparatively in both architectural styles. As a preliminary result, they found that the total amount of technical debt grew much faster in the Microservice-based system.

Bogner et al. \cite{Bogner2018a} surveyed 60 software professionals via an online questionnaire to assess maintainability assurance practices in industry as well as notable differences with service-oriented systems. The authors inquired information about the used processes, tools, and metrics to learn about treatments specific to such systems. Explicit and systematic techniques turned out to be beneficial for maintainability. Very few participants reported the usage of techniques to address existing issues related to architecture-level evolvability. Si\-milarly, barely any participants had established a systematic process for maintainability control. Since 67\% of participants neglected service-oriented particularities in the assurance, the study revealed a weak spot in industry practice that may impair the lifespan of service-based systems. 

In summary, the majority of mentioned studies does not focus on applied assurance techniques in industry. Bogner et al. are the only ones to do so in their survey. However, they also include service-based systems and say very little about participants' rationales. We therefore conducted our own qualitative Microservices study focused on industry evolvability assurance processes, techniques, and challenges.

\begin{table*}
    \centering
	\caption{Company and Participant Demographics}
	\label{table:demographics}
	\begin{tabular}{llrrlrr}
		Company ID & Company Domain & \# of Employees & Participant ID & Participant Role & Years of Experience & System ID\\
		\hline
		\hline
		C1 & Financial Services & 1 -- 25 & P1 & Developer & 6 & S1\\
		\hline
		\multirow{3}{*}{C2} & \multirow{3}{*}{Software \& IT Services} & \multirow{3}{*}{$>$100,000} & P2 & Lead Architect & 30 & S2\\
		 & & & P3 & Architect & 24 & S3\\
		 & & & P4 & Architect & 30 & S4\\
		\hline
		\multirow{2}{*}{C3} & \multirow{2}{*}{Software \& IT Services} & \multirow{2}{*}{26 -- 100} & P5 & Architect & 20 & \multirow{2}{*}{S5}\\
		 & & & P6 & Lead Developer & 8\\
		\hline
		\multirow{2}{*}{C4} & \multirow{2}{*}{Software \& IT Services} & \multirow{2}{*}{101 -- 1,000} & P7 & Architect & 9 & S6\\
		 & & & P8 & Architect & 17 & S7\\
		\hline
		C5 & Software \& IT Services & $>$100,000 & P9 & Lead Developer & 7 & S8\\
		\hline
		\multirow{3}{*}{C6} & \multirow{3}{*}{Tourism \& Travel} & \multirow{3}{*}{1,001 -- 5,000} & P10 & Developer & 9 & \multirow{2}{*}{S9}\\
		 & & & P11 & Data Engineer & 7\\
		 & & & P12 & Architect & 12 & S10\\
		\hline
		\multirow{2}{*}{C7} & \multirow{2}{*}{Logistics \& Public Transport} & \multirow{2}{*}{101 -- 1,000} & P13 & DevOps Engineer & 5 & S11\\
		 & & & P14 & Architect & 17 & S11a\\
		\hline
		C8 & Retail & 5,001 -- 10,000 & P15 & Lead Architect & 9 & S12\\
		\hline
		C9 & Software \& IT Services & 101 -- 1,000 & P16 & Architect & 18 & S13\\
		\hline
		C10 & Retail & 1,001 -- 5,000 & P17 & Architect & 22 & S14\\
		\hline
	\end{tabular}
\end{table*}

\section{Research Design}
We generally followed the five-step case study process as described by Runeson and Höst \cite{Runeson2009} to structure our research. The different steps are described in the following subsections.

\subsection{Study Design}
As a first step, we defined our research objective and related research questions. The primary goal of this study can be formulated in the following way:

\begin{center}
    \textit{
        Analyze the applied \textbf{evolvability assurance}\\
    	For the purpose of knowledge generation\\
    	With respect to common practices and challenges\\
    	From the viewpoint of software professionals\\
    	In the context of \textbf{Microservices} in industry\\
    }
\end{center}

We formulated three research questions to set a direction and scope for our study:

\noindent\textbf{RQ1:} How do software professionals structure the general evolvability assurance activities for Microservices and for what reasons?

\noindent\textbf{RQ2:} What tools, metrics, and patterns do software professionals use for assuring the evolvability of Microservices and with what rationales?

\noindent\textbf{RQ3:} How do software professionals perceive the quality of their Microservices and assurance processes and what parts are seen as challenging?

Since quantitative survey research with questionnaires would not be in-depth enough to cover participants' rationales, we chose a qualitative research approach. Qualitative methods analyze relationships between concepts and directly deal with identified complexity \cite{Seaman2008}. Results from such methods are therefore very rich and informative and can provide insights into the thought process of participants. We selected semi-structured interviews \cite{Seaman2008}\cite{Hove2005} as a concrete method. This left us with a basic agenda, but also allowed us to dynamically adapt our questions based on responses. For our interview participants, we defined the following requirements:

\begin{itemize}
    \item Significant professional experience (minimum of five years) and solid knowledge of service orientation
    \item Technical role (e.g. developer or architect) that at least sometimes writes code
    \item Recent participation in the development of a system with Microservices characteristics
\end{itemize}

We recruited participants via personal industry contacts of the research group and by attending industry meet-up groups on Microservices, where we approached companies from different domains and of different size. To ensure a base degree of heterogeneity within our population, we only allowed a maximum of three participants per company and if two participants worked on the same system, they needed to have different roles.

\subsection{Preparation for Data Collection}
Before conducting the interviews, we created several documents. We prepared an \textit{interview preamble} \cite{Runeson2009} that explained the interview process and relevant topics. To make participants familiar with the study, they received this document beforehand. The preamble also outlined ethical considerations like confidentiality, requested consent for audio recordings, and guaranteed that recordings and transcripts would not be published. As a second document, we created an \textit{interview guide} \cite{Seaman2008} that contained the most important questions grouped in thematic blocks. This guide helped us to scope and organize the semi-structured interviews and was used as a loose structure during the sessions. We did not share it with interviewees beforehand. Additional interview artifacts used as a reference for certain topics or questions, like e.g. an exemplary list of assurance tools, were collected in a slide set which was was also not shared with participants before the interviews. For the analysis, we created a preliminary set of \textit{coding labels} and a \textit{case characterization matrix} \cite{Seaman2008} containing the most important case attributes.

\begin{figure}[ht]
    \centering
    \includegraphics[keepaspectratio=true, width=0.51\textwidth]{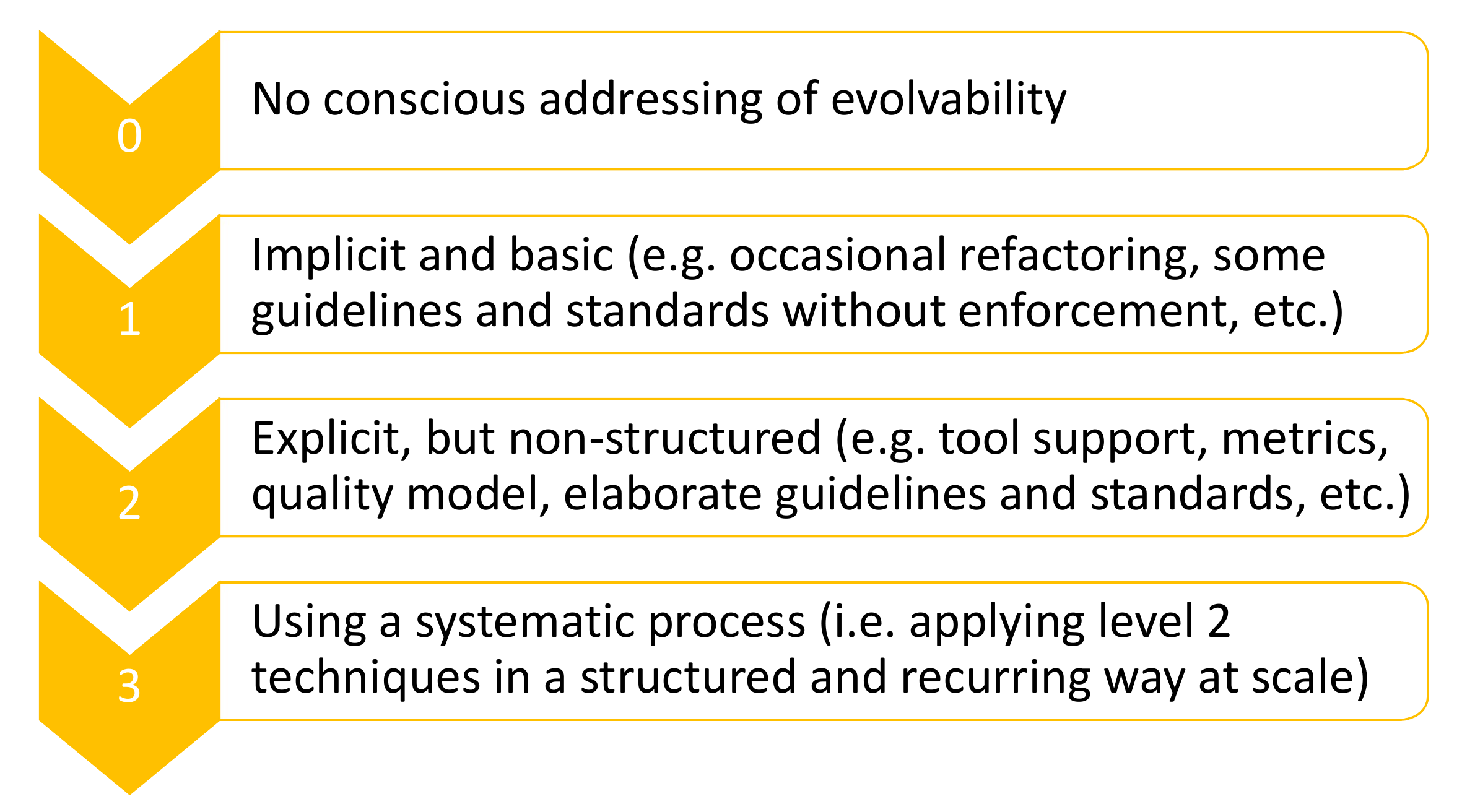}
    \caption{Evolvability Assurance Maturity Levels}
    \label{fig:maturity-levels}
\end{figure}

\subsection{Evidence Collection}
In total, we conducted 17 individual interviews (no group interviews). Six of these were performed face to face and 11 via remote communication software with screen sharing. All interviews except for an English one were conducted in German. Each participant agreed to a recording of the interviews, which took between $\sim$45 and $\sim$75 minutes. We loosely followed the structure of the interview guide and adapted based on how a participant reacted. As an initial ice breaker, participants were asked to describe their role and the system they worked on. Later on, the topic shifted to the evolvability of the system and potential symptoms of technical debt. The next thematic block was the concrete assurance process for the system. We presented a maturity model with four levels that ranged from implicit and basic to explicit and systematic (see Fig. \ref{fig:maturity-levels}). Participants were then asked to place themselves on the level that corresponded the most to their current assurance activities. From there, we discussed the details of the process such as techniques, tools, and metrics. Lastly, we asked questions about challenges and participants' satisfaction with the current process. The satisfaction and reflection questions relied on a five point Likert scale from -2 (very negative) to +2 (very positive) with 0 being the neutral center. After the interviews, we manually transcribed each audio recording to create a textual document. We then sent these documents to participants for review and final approval. During this review, interviewees were able to delete sensitive paragraphs or change statements of unclear or unintended meaning. The approved transcripts were then used for detailed qualitative content analysis.

\subsection{Data Analysis}
As a first step, we performed the coding of each transcript. Using the created preliminary set of codes, we assigned labels to relevant paragraphs. During this process, several new labels were created and already finished transcripts were revisited. Labels were also renamed, split, or merged as we acquired a more holistic understanding of the cases. These coding activities followed the \textit{constant comparison method} that is based on grounded theory \cite{Seaman2008}. After the coding of all transcripts, we analyzed the details and code relationships of each individual transcript. This activity resulted in a textual description for every case.

In the second step, we applied \textit{cross-case analysis} \cite{Seaman2008} to identify important generalizations and summaries between the cases. We used the coding system and the created case characterization matrix as well as \textit{tabulation} \cite{Runeson2009}. For each research question, important findings were extracted from the transcript and documented. During this process, we also refined the case characterization matrix. General trends and deviations were documented and later aggregated into results and take-aways. All interview documents and artifacts as well as the results of the analysis are available in our anonymous online repository\footnote{https://doi.org/10.5281/zenodo.2586916}.

\begin{table*}
    \centering
	\caption{General Case Characteristics: Assurance Maturity Level, Tools, Metrics, and Patterns}
	\label{table:cases}
	\begin{tabular}{
	    p{0.01\textwidth}
	    p{0.12\textwidth}
	    p{0.06\textwidth}
	    p{0.08\textwidth}
	    p{0.2\textwidth}
	    p{0.2\textwidth}
	    p{0.13\textwidth}
	}
		ID & System Purpose & \# of Services & Assurance Maturity Level (0-3) & Tools & Metrics & Patterns\\
		\hline
		\hline
		S1 & Derivatives management system (banking) & 9 & 2 & IDE linting & -- & Event-Driven Messaging, Service Registry\\
		\hline
		S2 & Freeway toll management system & 10 & 3 & SonarQube (FindBugs, Checkstyle, PMD) & Test coverage, cyclomatic complexity, clone coverage, \# of defects per service, \# of failed tests, \# of code smells, \# of endangered requirements & Event-Driven Messaging\\
		\hline
		S3 & Automotive problem management system & 10 & 1.5 & SonarQube (FindBugs, Checkstyle, PMD) & Test coverage, clone coverage, defect resolution time & Event-Driven Messaging, Strangler\\
		\hline
		S4 & Public transport sales system & $\sim$100 & 1.5 & SonarQube (FindBugs), VersionEye & \# of code smells, test coverage, \# of outdated dependencies & Event-Driven Messaging, Backends for Frontends, Consumer-Driven Contracts, Tolerant Reader\\
		\hline
		S5 & Business analytics \& data integration system & 6 & \makecell[tl]{P5: 1\\P6: 1.5} & SonarQube (FindBugs, Checkstyle, PMD) & Test coverage & Event-Driven Messaging\\
		\hline
		S6 & Automotive configuration management system & 60 & 3 & SonarQube (FindBugs, Checkstyle) & Test coverage, \# of failed tests, \# of code smells, cyclomatic complexity, clone coverage, LOC & Event-Driven Messaging\\
		\hline
		S7 & Retail online shop & $\sim$250 & 2.5 & SonarQube (FindBugs, PMD, Checkstyle), Cobertura, IDE linting, custom static analyzer for coding conventions & Test coverage, cyclomatic complexity, clone coverage, \# of rule violations, velocity & API Gateway\\
		\hline
		S8 & IT service monitoring platform & 9 & 2 & SonarQube (FindBugs), IDE linting & Cognitive complexity, \# of code smells, test coverage & Event-Driven Messaging\\
		\hline
		S9 & Hotel search engine & $\sim$10 & \makecell[tl]{P10: 1.5\\P11: 3} & \makecell[tl]{P10: Checkstyle, IDE linting\\P11: SonarQube (FindBugs,\\Checkstyle)} & \makecell[tl]{P10: Defect resolution time,\\burndown\\P11: \# of code smells,\\\# of rule violations} & Event-Driven Messaging, Request-Reaction\\
		\hline
		S10 & Hotel management suite & 20 & 2 & SonarQube, IDE linting & Test coverage, \# of endangered usage scenarios & Self-Contained Systems, Backends for Frontends\\
		\hline
		S11 & Public transport management suite (S11a: human resource management part) & \makecell[tl]{10\\products} & \makecell[tl]{P13: 1.5\\P14: 2.5} & \makecell[tl]{P13: SonarQube (FindBugs,\\Checkstyle)\\P14: FindBugs, PMD, Cobertura,\\custom tool for architectural\\conformance checking} & \makecell[tl]{P13: \# of code smells\\P14: \# of architectural violations,\\\# of rule violations} & --\\
		\hline
		S12 & Retail online shop & $\sim$45 & 2.5 & SonarQube (FindBugs, Checkstyle, PMD), IDE linting, Codecov, Cobertura & Test coverage, cyclomatic complexity, \# of code smells, \# of rule violations & Event-Driven Messaging, Consumer-Driven Contracts\\
		\hline
		S13 & Automotive end-user services mgmt. system & 7 & 2 & SonarQube (FindBugs, Checkstyle, PMD) & Test coverage, \# of code smells, \# of rule violations & Event-Driven Messaging\\
		\hline
		S14 & Retail online shop & $\sim$175 & 2 & SonarQube (FindBugs, Checkstyle, PMD), Structure101, Codecov, Cobertura, Codacy & Test coverage, CI/CD pipeline duration, LOC & Event-Driven Messaging, Self-Contained Systems, Event Sourcing\\
		\hline
	\end{tabular}
\end{table*}

\section{Interview Case Characteristics}
Our interviewees were from 10 different companies (C1--C10) of different sizes and domains (see Table \ref{table:demographics}). Half of these were software \& IT services companies that mostly developed systems for external customers. The companies from other domains always had an internal system owner. Every participant was located in Germany, even though some companies had sites in several European countries or even globally. From our 17 participants (P1--P17), 11 stated architect as their role while four were developers. The remaining two roles were data engineer and DevOps engineer. All participants possessed a minimum of five years of professional experience, with a median of 12 and a mean of 14.7 years. Altogether, we discussed 14 systems (S1--S14) and their evolvability assurance processes, where in three cases, two participants talked about the same system (S5, S9, S11).

\begin{figure*}[ht]
    \centering
    \includegraphics[keepaspectratio=true, width=1\textwidth]{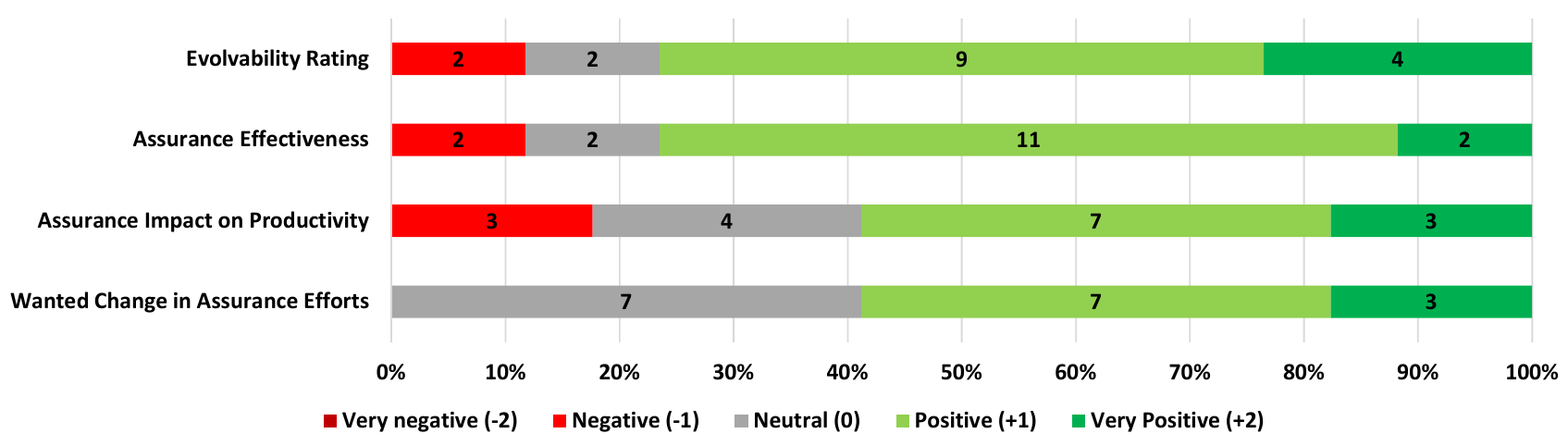}
    \caption{Aggregated Evolvability Assurance Reflections of our 17 Participants}
    \label{fig:reflections}
\end{figure*}

For the sake of brevity, this publication only contains the aggregated results. We provide a detailed description of every case in our online repository$^1$. The descriptions include general information about the system, the details of the evolvability assurance process, and lastly reflections and challenges. Table \ref{table:cases} summarizes the self-assessed assurance maturity levels as well as the usage of tools, metrics, and patterns to analyze and improve evolvability. Fig.~\ref{fig:reflections} presents the summary of participants' reflections of their assurance processes. Lastly, the most important challenges reported by participants are listed in Fig.~\ref{fig:challenges}.

\section{Results and Discussion}
By analyzing and comparing the individual cases, we identified several trends or common relationships. These generalizations, summaries, or notable deviations from common assumptions are presented in the following subsections that correspond to our three research questions.

\subsection{Assurance Processes (RQ1)}
The intention of RQ1 was to find out what general techniques participants used to assure the evolvability of Microservices, how they structured these activities, and what participants' rationales for these decisions were. Since one Microservices characteristic is the decentralization of control and management, we also wanted to analyze how much central governance was applied.

In general, every analyzed assurance process had some degree of explicit and conscious addressing of evolvability, even though the sophistication and extensiveness of the applied techniques varied greatly. When looking at the larger systems, there were \textbf{two different approaches} for assuring evolvability: very decentralized with very autonomous teams (e.g. S9, S10, S12, S14) vs. centralized governance for macroarchitecture, technologies, and assurance combined with a varying degree of team autonomy for microarchitecture (e.g. S2, S3, S4, S7, S13). The latter kind was usually applied for systems that were built for external customers and that exhibited some project characteristics.

In the decentralized variant, the internal system was managed in a continuous product development mode, which created quality awareness by making people responsible and simultaneously empowering them. This variant is more in line with the Microservices and DevOps principle \enquote{you build it, you run it}. Techniques in this variant also were by no means basic or implicit. Even though teams were allowed to choose their own assurance activities, they usually created a more or less structured processes that did not depend on external governance. This was hard to replicate for IT service providers that often did not operate the systems themselves (S2, S3, S4, S7, S13) and had to coordinate with external customers or even other contractors (S4, S7). Therefore, they relied more on central governance. Architect P4 described it as follows: \enquote{\textit{In our case, the main challenge is to convince 300 people to move in the same direction. For that, we created a very large number of guidelines and rules for service creation.}}

Such \textbf{guidelines, principles, or standardizations} were nonetheless seen as important parts of the process in both variants. These coding labels were among the most frequent ones. Nearly all participants reported their usage in various areas such as architectural principles, rules for service communication, skeleton projects, style guides, cross-cutting concerns like logging or authentication, candidate technologies, or Docker images. The degree of enforcement varied between companies and was usually higher within the centralized variant. In the decentralized variant, pragmatism and simplicity were often more important than the strict adherence to rules and were reported as key principles (\enquote{KISS}) by several participants (P5, P8, P10, P12, P15, P17).

To make assurance activities more efficient and objective, all participants saw \textbf{automation and tool support} as useful, albeit with varying enthusiasm. Several participants reported the integration of quality analysis tools into the CI/CD pipeline (P2, P3, P8, P9, P11, P14, P15, P17). This was often combined with quality gates that could prevent merging or deployment. For P17, the pipeline's execution time was very important: only tools that were absolutely necessary should therefore be integrated. Additionally, several participants advocated for a sensible usage of quality gates. P11 was frustrated with how difficult it would be to get a passing merge request due to the strict rules. Similarly, P10 mentioned that strict quality gates could hinder the deployment of important production bug fixes. P7's team did not use any quality gates because of continuous experimentation and prototyping. Lastly, P2's team circumvented some of these issues by applying quality gates only for releases and not for merge or pull requests.

Nearly all participants agreed that test automation was an important part for the assurance process of Microservices. While unit tests were very common, several participants also reported automated end-to-end tests for the integration of Microservices and stressed their importance (P2, P3, P7, P12, P16). Some teams also had more elaborate strategies that linked tests to requirements (S2) or usage scenarios (S10). Participants with only unit tests (P5, P6, P9) or barely any tests (P10) also saw the importance to bring their test automation to a higher level.

Despite the reported importance of automation and tool support, several participants also highlighted the usefulness of \textbf{manual assurance activities}. Code reviews were seen as an important practice to increase code quality and to share knowledge within the team (P1, P3, P4, P5, P7, P8, P10, P11). Pair programming was used for the same reasons by two participants and the downside of additional man-hours was willingly accepted (P8, P15). Lastly, refactoring was highly valued and some participants also explicitly mentioned the use of boy scouting during feature implementations (P11, P15) to efficiently increase code quality over time.

Even though some participants were proponents of concise documentation or architectural decision records within the source code repository (P10, P11), several systems relied on more elaborate \textbf{architecture and service documentation} in a system like Confluence or SharePoint (S1, S4, S5, S6, S8). Common types of documentation were system architecture, service dependencies and contracts between teams, service functionality and API descriptions, reference architectures and service blueprints, design rationales, or architectural principles and guidelines. For IT service providers, parts of this documentation was also used to communicate with the customer. Lastly, only P7 and P14 reported the conscious tracking of identified technical debt items for later debt management. P7's teams held an explicit meeting every two weeks, where the most important technical debt items were discussed and their prioritization was decided.

\subsection{Tools, Metrics, and Patterns (RQ2)}
Our second research question targeted the application of and rationale for tools, metrics, and design patterns. Automation and tool support is an often cited Microservices characteristic and seen as necessary to manage a large number of small components. We were also interested if participants used tools and metrics specifically designed for service orientation. Lastly, we wanted to explore the usage of design patterns for evolvability construction, since there is a large body of patterns for Service-Oriented Architecture (SOA) and more recently also for Microservices.

While the usage of over a dozen different \textbf{tools for evolvability assurance} was reported, 14 of 17 participants named SonarQube as a central tool that was usually integrated into the CI/CD pipeline. Since P1 and P10 planned to introduce it soon, P14 remained the only participant that would not use SonarQube in the foreseeable future. Reported reasons for its popularity were the OpenSource license, the easy installation, plugin availability, and configurability. In Java-focused systems, SonarQube was often extended with tools like FindBugs, Checkstyle, and PMD. Additionally, specialized tools for test coverage like Cobertura (P8, P14, P15, P17), Codecov (P15, P17), or Codacy (P17) were used. For a basic degree of local and immediate quality assurance, IDE linting via e.g. TSLint, ESLint, and PHPLint was reported by some participants (P1, P8, P10, P12, P15).

With respect to \textbf{metrics}, 10 of the 17 participants reported the usage of test coverage, even though some perceived this metrics as less important than others and were very aware of possible quality differences with automated tests. P12 termed it as follows: \enquote{\textit{Even I could fake the coverage for two classes you give me in like five minutes. You can write a test that brings coverage to about 60\%, but actually it covers like 2\%.}} Some participants also focused on additional metrics for testing and functional correctness like the number of failed tests over time (P2, P7), the number of defects per service (P2), or the number of endangered requirements (P2) or usage scenarios (P12). Most SonarQube users also payed attention to standard findings like code smells, code duplication, and cognitive or cyclomatic complexity. Participants with rule-based tools like FindBugs, Checkstyle, or other linters used the number of rule violations as a simple metric that had to be zero.

Overall, most applied metrics were focused on source code quality, even though their effectiveness for the whole system was seen as controversial by a few participants (P8, P17). P17 described it as follows: \enquote{\textit{Most of these metrics relate to a single project, which is very useful when I have a monolith with a million LOC. However, if I have a service with 1000 LOC which code base is separated from all other 150 Microservices, most of these metrics lose their importance.}} With respect to productivity metrics, some interviewees reported the usage of defect resolution time (P3, P10), velocity (P8), sprint burndown (P10), or deployment duration (P17). These were important for them to control and manage software evolution.

\begin{figure*}[ht]
    \centering
    \includegraphics[keepaspectratio=true, width=1\textwidth]{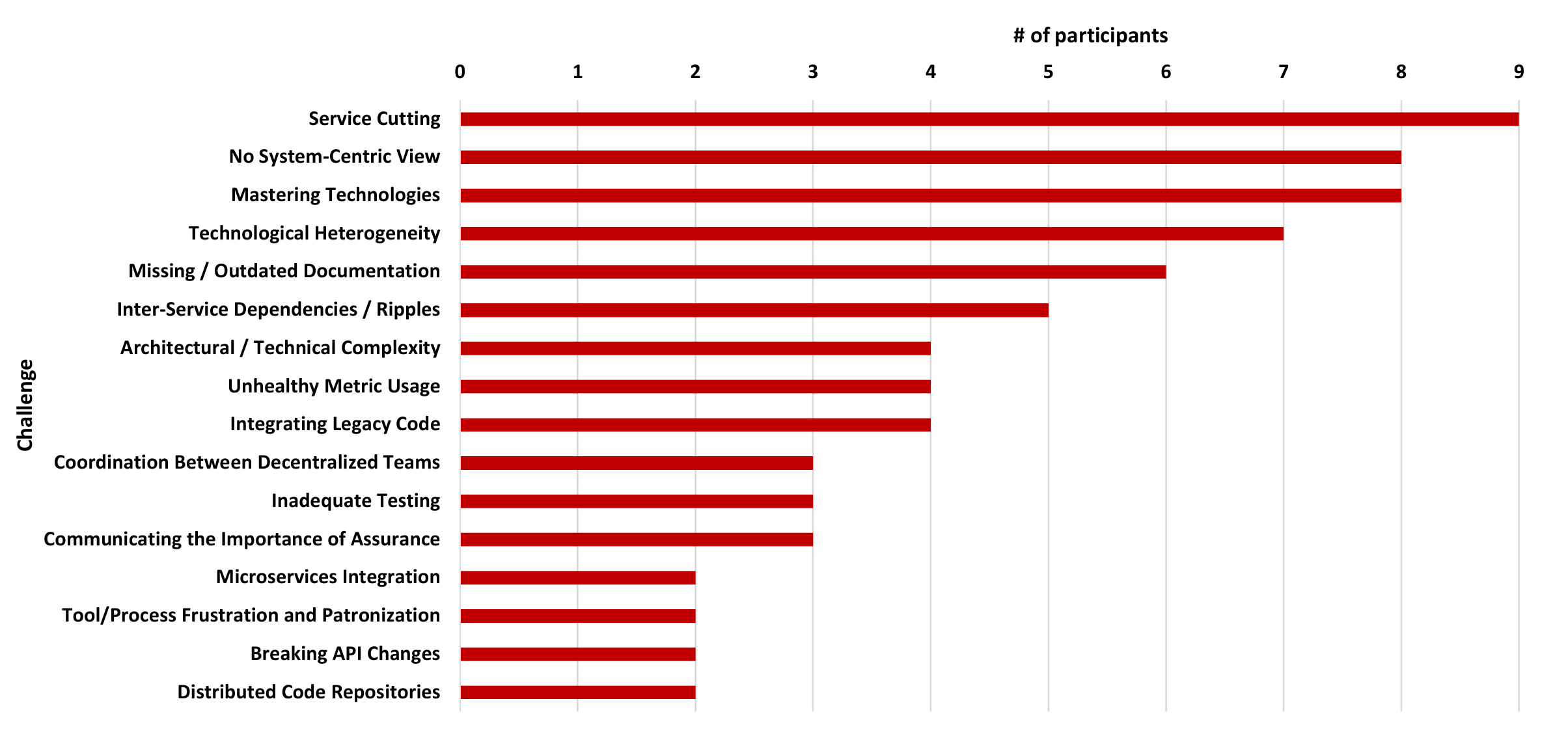}
    \caption{Most Important Challenges for the Evolvability of Microservices as Reported by our 17 participants}
    \label{fig:challenges}
\end{figure*}

While architecture-related topics like Microservices dependencies were very prevalent during our interviews, participants generally did not apply \textbf{architecture-level tools and metrics}. P14's team used a custom tool for architectural conformance checking in the monolithic code base for a sub product of S11 and P17 reported the intermittent usage of Structure101 for a larger subsystem that also consisted of one code base. Apart from that, tools or metrics were exclusively focused on code quality with a local view for a single service. No (semi-)automatic efforts were mentioned to evaluate the architecture of a Microservice-based System.

Likewise, no participant reported the usage of a tool or metric specifically designed for service orientation. When we explicitly asked about service-oriented metrics like the coupling of a service, the cohesion of a service interface, or the number of operations in a service interface, several participants indicated that these sounded interesting and useful (P1, P5, P6, P7, P8, P15). Some interviewees also noted that the underlying principles of these metrics were important guidelines in the architecture and design phase (P7, P8, P10, P15, P17). They tried to manually respect these principles during e.g. service cutting, even though they currently had no concrete measurements in place to validate them.

Another common theme in this area was the \textbf{healthy and non-patronizing usage of tools and metrics}, which should be respected when developing Microservices in decentralized and fairly autonomous teams. As already mentioned, several participants voiced reservations against test coverage (P10, P12, P15). P14 also warned that a strict metric focus would pose the danger that people optimized for measurements instead of fixing the underlying problems. Moreover, P15 perceived it as difficult to interpret measurements of a single service without a point of reference or a system-wide average (P15). P17 advocated for a sparse usage of tools, because too many metrics could not be analyzed by developers and their collection could slow down the deployment pipeline. Only tools that would support the analysis of current problems should be kept. Lastly, P8 and P14 highlighted the agile principle of \enquote{individuals and interactions over processes and tools}: the usage of tools and metrics should support developers in their daily work and not be a frustrating and alienating experience for them.

We also analyzed the usage of \textbf{service-oriented design patterns} as conscious means to increase evolvability. In general, we did not find a widespread usage of them. Most common was \textit{Event-Driven Messaging} that was partially applied in 11 of the 14 cases. While several participants stated that the pattern was used to decouple services, another intention was to implement reliable asynchronous and long-running communication. The pattern was sometimes paired with \textit{Request-Reaction} (P10, P16). Apart from messaging, most participants applied activity patterns like \textit{Service Refactoring}, \textit{Service Decomposition}, and \textit{Service Normalization}. In line with the philosophy of evolutionary design, Microservices were frequently split and merged.

Other patterns were used sporadically. P12 and P17 applied the \textit{Self-Contained System} paradigm to achieve vertical isolation between subsystems. In a migration context, P3 and P16 reported the usage of the \textit{Strangler} pattern to extend an existing monolith with new Microservices until its final replacement. To place an intermediary between service consumers and producers, P4 and P12 implemented the \textit{Backends for Frontends} pattern that would also prevent too many concurrent long-running HTTP requests. Similarly, P8 chose the \textit{API Gateway} pattern which also brought benefits for security. The patterns \textit{Consumer-Driven Contracts} (P4, P15) and \textit{Tolerant Reader} (P4) were applied to make service interface evolution more robust and to prepare consumers for future changes. Lastly, P1 was the only participant to explicitly report the usage of the \textit{Service Registry} pattern for dynamic service discovery, even though some participants may have used similar functionality via Kubernetes.

\subsection{Evolvability Reflections and Challenges (RQ3)}
With RQ3, we wanted to analyze participants' perception of the general evolution qualities of their Microservice-based systems as well as their satisfaction with their current assurance processes (see Fig. \ref{fig:reflections}). We also tried to summarize what participants experienced as the most important challenges for the evolvability of their Microservices (see Fig. \ref{fig:challenges}).

Generally, our interviewees perceived the \textbf{evolvability of their Microservices} as positive (mean: +0.88, median: +1), especially in cases with a migration context where a monolith had been rewritten. Only two participants chose a negative rating (-1). Architect P4 saw the high degree of technological heterogeneity and the very different service granularity as threatening for the large project, especially once S4 would be handed over to the smaller maintenance team. Data engineer P11 described the chosen service cuts as inefficient and politically motivated and worried about significant issues with the consistency of the data model as well as the inaccessibility of code due to distributed repositories.

As for more specific quality attributes, the analyzability of individual services would be much improved (P1, P8, P10, P16, P17), even though grasping and understanding the whole system would be difficult (P7, P8, P11, P17). When compared to most monoliths, the modularity of Microservices would make it very convenient to change or add functionality (P1, P3, P6, P9, P12, P17) and would also allow to efficiently scale-out the development with multiple teams (P7, P16). Even though reuse is usually a theme more common in SOA, several participants reported a positive reusability of their Microservices (P1, P3, P7, P8, P10, P17). To reduce coupling, some participants avoided the sharing of non-OpenSource code between services via duplication (P7, P10, P15, P17). Others tried to consciously increase reuse via shared libraries (P3, P9) or by slightly generalizing service interfaces (P1). Lastly, participants reported that individual services would be easy to test (P3, P7, P10, P13, P15) and to replace (P2, P15, P17).

Since most systems were fairly young or even still in the process of being migrated, individual services were usually of a good quality. Basic symptoms of technical debt or bad code quality were rarely seen as an issue, especially since a single service would be easy to replace. However, \textbf{problems related to architecture and the data model} were reported as serious threats for long-term evolvability (P3, P7, P11, P13, P15). This was sometimes exacerbated, because coordination between autonomous teams would be difficult (P4, P10, P11, P15). Moreover, finding the appropriate service granularity was a prevalent theme and service cutting was by far named as the most challenging activity that was also associated with frequent refactoring (P2, P3, P4, P6, P7, P9, P11, P12, P15). Harmful inter-service dependencies sometimes led to ripple effects on changes (P3, P5, P9, P11, P15) which made adding or changing functionality slower and more error-prone. Breaking API changes caused similar effects for service consumers (P1) or automated tests (P2). Participants did not use any tools to support service decomposition or metrics to evaluate the quality of the chosen cuts, e.g. via coupling or cohesion. P2 described it as follows: \enquote{\textit{In my opinion, there are no useful tools to split up a monolith. It’s always a very difficult manual activity. You can use something like Domain-Driven Design, but that’s just a methodology which doesn’t give you a concrete solution.}}

Participants were divided when it came to \textbf{technological heterogeneity}. In very decentralized environments, it was generally perceived as overall beneficial (P10, P15, P17), as it would allow choosing the best solution for problems at hand, broaden developers' experience and skills, and make a company a more attractive employer. Other participants perceived it as potentially dangerous and wished for a more sensible handling of technology hypes (P3, P4, P12, P16). Similarly, the mix of legacy and modern service technology would sometimes pose additional problems (P2, P11, P13, P14), like in the case of S9 where additional tooling was necessary to integrate legacy PHP components. Most participants also noted that significant efforts had to be spent on mastering new Microservices and DevOps technologies and it would be problematic to find skilled developers (P1, P3, P5, P6, P9, P10, P12, P13, P16). Overall, participants were very aware of the human factors of evolvability and sometimes even saw them as more challenging as technical ones. Knowledge exchange between teams was therefore a high priority for some interviewees (P10, P13, P15).

Concerning participants' \textbf{reflection of their assurance processes} (see also Fig. \ref{fig:reflections}), most saw the effectiveness of their assurance activities (mean: +0.76, median: +1) as well as overall impact on productivity (mean: +0.59, median: +1) as positive. Only three interviewees (P9, P11, P12) reported that activities would hinder development efficiency (-1) and would sometimes slow down feature development. Moreover, participants generally wanted to invest more effort for the assurance (mean: +0.76, median: +1) and try out new techniques or metrics. No one reported that he would like to reduce efforts. 

However, the influence of Microservices on the assurance process was seen as controversial. While testing a single service would be easy, integration testing would be more complex because of an additional layer (P2, P3, P13). This would be especially critical, if Microservices were developed in independence for a long time and integrated at a later stage. Furthermore, root cause analysis of issues would be more complex in such a highly distributed system (P3, P11). A very commonly named concern was that keeping a system-centric quality view and assessing the macroarchitecture would be much more difficult (P4, P6, P7, P8, P10, P11, P15, P17), which P8 described as follows: \enquote{\textit{I’d say we are pretty good when it comes to assuring the evolvability of single services. However, we have a lot of catching up to do for everything that crosses product or service boundaries.}} Distributed code repositories and autonomous teams would make the access to code as well as static analysis more complicated. It would also be hard to compare metrics between services and relate them to system-wide averages (P8, P11, P17).

Nonetheless, participants also named positive factors. Small services would not only be easy to replace, people would also be much more motivated to fix a small number of issues for a project (P15, P17), which P15 described as follows: \enquote{\textit{In a monolith with 100.000 FindBugs warnings, you are completely demotivated to even fix a single one of those. In a Microservice with 100 warnings, you just get to work and remove them.}} If adopted correctly, Microservices would also bring a cultural change with respect to quality awareness and responsibility (P10, P15, P17). P17 highlighted the importance of continuous product development in this regard: \enquote{\textit{If you work in a project mode, evolvability assurance usually annoys you, because you have short-term goals and want to finish the project. In a product mode, the team knows that they sabotage their system’s evolvability in the long run, if they take too many short-cuts.}} Lastly, P15 noted that while they were relatively satisfied with their current evolvability assurance activities, they did not really invest much efforts into researching and designing a fitting evolvability assurance strategy for the future. Finding out which approaches, tools, and metrics worked best for their Microservices could be a vital advantage in the long-term.

\section{Threats to Validity}
A number of possible limitations have to be mentioned for this research. Concerning \textbf{internal validity}, it could be possible that interviewees were not completely honest during their answers, which is a common problem in survey- or interview-based studies. Nonetheless, we think that the risk for this study is rather low. Most participants were quite comfortable to describe negative aspects of their systems and the discussed topics were only of low to medium sensitivity. Furthermore, confidentiality and anonymity was provided. As a second threat, participants could have provided incorrect answers, because they misunderstood questions or concepts. To limit this risk, important concepts were defined before the corresponding topics. Moreover, we posed additional clarifying questions if participants used terms of unclear meaning. Some interviewees also asked questions themselves, if questions or terms were not fully clear to them. In some cases, our analysis relied on the subjective opinion of participants, e.g. how they rated the evolvability of their system. Since participants could interpret such ratings very differently, detailed comparisons between systems may be difficult. Moreover, we only relied on one single opinion for most systems and people may have different opinions of systems, as is visible in the case of S9.

To increase interpretation validity and reduce researcher bias, every transcript was proofread by both moderators. Additionally, we strongly encouraged our participants to adjust wrong or unclear statements in the transcripts. Nevertheless, there still remains the small possibility that some of the more figurative paragraphs were misinterpreted. Overall, the largest threat in qualitative studies is that summarizing and aggregating results depends heavily on the subjective interpretation of researchers. We tried to limit this bias by relying on a precise coding system and by critically discussing all results between the moderators.

Concerning \textbf{external validity}, no distributions can be generalized from our 17 interviews, e.g. the industry usage of SonarQube for Microservices. Because this is a qualitative study, the center of this research are the intentions and opinions of participants as well as the relationships between concepts. An additional threat for generalizability could be that our participants were exclusively based in Germany, even though we included European and international companies. We also tried to achieve diversity with respect to company domains and sizes, but had nine participants from software and IT services companies ($\sim$52\%). Lastly, possible selection bias could only be reduced in the case of C2 where we applied a random sampling to pick three out of seven candidates.

\section{Conclusion}
We conducted 17 interviews with software professionals from 10 companies and talked with them about the evolvability assurance for 14 different Microservice-based systems. We found that systems developed for an external customer in a project-like manner generally relied more on central governance for the assurance. Continuous product development of internal systems exhibited more decentralization and team autonomy for the assurance process. Guidelines like architectural principles or rules for service communication were seen as important to provide a consistent basis for evolvability and to frame an otherwise diverse technology usage. While participants saw automation and tool support as important for evolvability assurance, they still relied a lot on manual activities like code review or boy scouting. Tool and metric usage was very focused on source code quality. No architectural tools and metrics were applied, even though most reported challenges and issues like service cutting were related to software architecture. This may indicate the importance of Architectural Technical Debt (ATD) management for Microservices. Likewise, no tools or metrics specifically designed for service orientation were used, even though most participants stated the significance of their underlying design principles. Participants generally perceived their fairly new Microservices as decently evolvable, even though things like distributed code repositories and difficult macroarchitectural assessment would make the assurance process more complex.

Future work that covers the areas of maintenance, evolution, and technical debt in the context of Microservices should take these findings and industry sentiments into account. In particular, academia can support industry with methods, metrics, or tools that aid macroarchitectural assessment of Microservices or provide a more system-centric view. We perceived support for service cutting activities and metrics to evaluate service granularity, coupling, or cohesion as concrete gaps that could save industry a lot of refactoring efforts. Finally, issues in the area of human evolvability factors with Microservices like the handling of hyped technologies as well as coordinating and exchanging knowledge between decentralized teams were described as important by participants.


\bibliographystyle{IEEEtran}
\bibliography{references}

\begin{thebibliography}{10}
\providecommand{\url}[1]{#1}
\csname url@samestyle\endcsname
\providecommand{\newblock}{\relax}
\providecommand{\bibinfo}[2]{#2}
\providecommand{\BIBentrySTDinterwordspacing}{\spaceskip=0pt\relax}
\providecommand{\BIBentryALTinterwordstretchfactor}{4}
\providecommand{\BIBentryALTinterwordspacing}{\spaceskip=\fontdimen2\font plus
\BIBentryALTinterwordstretchfactor\fontdimen3\font minus
  \fontdimen4\font\relax}
\providecommand{\BIBforeignlanguage}[2]{{%
\expandafter\ifx\csname l@#1\endcsname\relax
\typeout{** WARNING: IEEEtran.bst: No hyphenation pattern has been}%
\typeout{** loaded for the language `#1'. Using the pattern for}%
\typeout{** the default language instead.}%
\else
\language=\csname l@#1\endcsname
\fi
#2}}
\providecommand{\BIBdecl}{\relax}
\BIBdecl

\bibitem{Avgeriou2016}
P.~Avgeriou, P.~Kruchten, I.~Ozkaya, C.~Seaman, and C.~Seaman, ``{Managing
  Technical Debt in Software Engineering},'' \emph{Dagstuhl Reports}, vol.~6,
  2016.

\bibitem{Rowe1998}
D.~Rowe, J.~Leaney, and D.~Lowe, ``{Defining Systems Architecture Evolvability
  - a taxonomy of change},'' \emph{International Conference and Workshop:
  Engineering of Computer-Based Systems}, 1998.

\bibitem{Bode2009}
S.~Bode, ``{On the Role of Evolvability for Architectural Design},'' in
  \emph{Informatik 2009: Im Focus das Leben, Beitr{\"{a}}ge der 39.
  Jahrestagung der Gesellschaft f{\"{u}}r Informatik e.V. (GI)}, ser. LNI, vol.
  154.\hskip 1em plus 0.5em minus 0.4em\relax GI, 2009.

\bibitem{Wagner2013}
S.~Wagner, \emph{{Software Product Quality Control}}.\hskip 1em plus 0.5em
  minus 0.4em\relax Berlin, Heidelberg: Springer Berlin Heidelberg, 2013.

\bibitem{Newman:15:MS}
S.~Newman, \emph{{Building Microservices: Designing Fine-Grained Systems}},
  1st~ed.\hskip 1em plus 0.5em minus 0.4em\relax O'Reilly Media, 2015.

\bibitem{Fowler2015}
M.~Fowler, ``{Microservices Resource Guide},'' 2015.

\bibitem{Fritzsch2019}
J.~Fritzsch, J.~Bogner, A.~Zimmermann, and S.~Wagner, ``{From Monolith to
  Microservices: A Classification of Refactoring Approaches},'' in
  \emph{Software Engineering Aspects of Continuous Development and New
  Paradigms of Software Production and Deployment}, J.-M. Bruel, M.~Mazzara,
  and B.~Meyer, Eds.\hskip 1em plus 0.5em minus 0.4em\relax Toulouse, France:
  Springer, 2019.

\bibitem{Esparrachiari2018}
S.~Esparrachiari, T.~Reilly, and A.~Rentz, ``{Tracking and Controlling
  Microservice Dependencies},'' \emph{ACM Queue}, vol.~16, 2018.

\bibitem{Sampaio2017}
A.~R. Sampaio, H.~Kadiyala, B.~Hu, J.~Steinbacher, T.~Erwin, N.~Rosa,
  I.~Beschastnikh, and J.~Rubin, ``{Supporting Microservice Evolution},'' in
  \emph{2017 IEEE International Conference on Software Maintenance and
  Evolution (ICSME)}.\hskip 1em plus 0.5em minus 0.4em\relax IEEE, 2017.

\bibitem{Bogner2017}
J.~Bogner, S.~Wagner, and A.~Zimmermann, ``{Automatically measuring the
  maintainability of service- and microservice-based systems},'' in
  \emph{Proceedings of the 27th International Workshop on Software Measurement
  and 12th International Conference on Software Process and Product Measurement
  on - IWSM Mensura '17}.\hskip 1em plus 0.5em minus 0.4em\relax New York, New
  York, USA: ACM Press, 2017.

\bibitem{Schermann2016}
G.~Schermann, J.~Cito, and P.~Leitner, ``{All the Services Large and Micro:
  Revisiting Industrial Practice in Services Computing},'' in \emph{Lecture
  Notes in Computer Science}, ser. LNCS.\hskip 1em plus 0.5em minus 0.4em\relax
  Berlin, Heidelberg: Springer Berlin Heidelberg, 2016, vol. 9586.

\bibitem{Baskarada2018}
S.~Ba{\v{s}}karada, V.~Nguyen, and A.~Koronios, ``{Architecting Microservices:
  Practical Opportunities and Challenges},'' \emph{Journal of Computer
  Information Systems}, vol.~00, 2018.

\bibitem{Ghofrani2018}
J.~Ghofrani and D.~L{\"{u}}bke, ``{Challenges of Microservices Architecture: A
  Survey on the State of the Practice},'' in \emph{10th Central European
  Workshop on Services and their Composition (ZEUS)}, vol. 10th.\hskip 1em plus
  0.5em minus 0.4em\relax Dresden, Germany: CEUR-WS.org, 2018.

\bibitem{Haselbock2018}
S.~Haselbock, R.~Weinreich, and G.~Buchgeher, ``{An Expert Interview Study on
  Areas of Microservice Design},'' in \emph{2018 IEEE 11th Conference on
  Service-Oriented Computing and Applications (SOCA)}.\hskip 1em plus 0.5em
  minus 0.4em\relax IEEE, 2018.

\bibitem{Carrasco2018}
A.~Carrasco, B.~van Bladel, and S.~Demeyer, ``{Migrating towards microservices:
  migration and architecture smells},'' in \emph{Proceedings of the 2nd
  International Workshop on Refactoring - IWoR 2018}.\hskip 1em plus 0.5em
  minus 0.4em\relax New York, New York, USA: ACM Press, 2018.

\bibitem{Taibi2018}
D.~Taibi and V.~Lenarduzzi, ``{On the Definition of Microservice Bad Smells},''
  \emph{IEEE Software}, vol.~35, 2018.

\bibitem{Lenarduzzi2018}
V.~Lenarduzzi and D.~Taibi, ``{Microservices, Continuous Architecture, and
  Technical Debt Interest: An Empirical Study},'' in \emph{Euromicro SEAA},
  Prague, Czech Republic, 2018.

\bibitem{Bogner2018a}
J.~Bogner, J.~Fritzsch, S.~Wagner, and A.~Zimmermann, ``{Limiting Technical
  Debt with Maintainability Assurance – An Industry Survey on Used Techniques
  and Differences with Service- and Microservice-Based Systems},'' in
  \emph{Proceedings of the 2018 International Conference on Technical Debt -
  TechDebt '18}.\hskip 1em plus 0.5em minus 0.4em\relax New York, New York,
  USA: ACM Press, 2018.

\bibitem{Runeson2009}
P.~Runeson and M.~H{\"{o}}st, ``{Guidelines for conducting and reporting case
  study research in software engineering},'' \emph{Empirical Software
  Engineering}, vol.~14, 2009.

\bibitem{Seaman2008}
C.~B. Seaman, ``{Qualitative Methods},'' in \emph{Guide to Advanced Empirical
  Software Engineering}.\hskip 1em plus 0.5em minus 0.4em\relax London:
  Springer London, 2008.

\bibitem{Hove2005}
S.~Hove and B.~Anda, ``{Experiences from Conducting Semi-structured Interviews
  in Empirical Software Engineering Research},'' in \emph{11th IEEE
  International Software Metrics Symposium (METRICS'05)}.\hskip 1em plus 0.5em
  minus 0.4em\relax IEEE, 2005.

\end{thebibliography}

\end{document}